\documentclass[aps,prc,showpacs,preprint,preprintnumbers,
showkeys,tighten,floatfix,nofootinbib]{revtex4}
\usepackage{amsmath,amssymb,amsfonts}
\usepackage{graphicx}
\usepackage{subfigure}
\usepackage{color}
\allowdisplaybreaks
\def \Slash{\slash \!\!\!}
\def \del{\partial}

\begin{document}

\title{A MODIFIED HARD THERMAL LOOP PERTURBATION THEORY}

\author{Najmul Haque and Munshi G. Mustafa}

\affiliation{Theory Division, Saha Institute of Nuclear Physics, 
          1/AF, Bidhannagar, Kolkata 700 064, India.}

\begin{abstract}
{Based on the external perturbation that disturbs the system only slightly
from its equilibrium position we make the Taylor expansion of the pressure
of a quark gas. It turns out that the first term was used in the literature
to construct a Hard Thermal Loop perturbation theory (HTLpt) within the 
variation principle of the lowest order of the thermal mass parameter. 
Various thermodynamic quantities  
within the 1-loop HTLpt encountered overcounting of the leading order (LO) 
contribution and also required a separation scale for soft and hard momenta. 
Using same variational principle we reconstruct the HTLpt at the first 
derivative level of the pressure that takes into account the effect of 
the variation of the external source through the conserved density fluctuation. This modification markedly improves those quantities 
in 1-loop HTLpt in a simple way instead of pushing the calculation to 
a considerably more complicated 2-loop HTLpt.  Moreover, the results also
agree with those obtained in the  2-loop approximately self-consistent 
$\Phi$-derivable Hard Thermal Loop resummation. We also discuss how this 
formalism can be extended for the higher order contributions }  
\end{abstract}

\pacs{12.38.Cy, 12.38.Mh, 11.10.Wx}
\keywords{Quark-Gluon Plasma, Hard Thermal Loop Approximation,
Quark Number Susceptibility}
\maketitle

\section{Introduction}
The HTL resummation developed by Braaten and Pisarski~\cite{braaten} 
has been used to calculate various thermodynamic quantities in the 
literature based on two methods. The 2-loop approximately $\Phi$-derivable 
approach was developed by Blaizot {\it et al}~\cite{blaizot1},
which produces a correct LO and plasmon effects for thermodynamic 
quantities (e.g, entropy density, number density etc.) 
and also for quark number susceptibility (QNS)~\cite{blaizot}. On the 
other hand the HTLpt using 
variational principle through the lowest order of the thermal mass
parameter was developed for pressure at the 1-loop level by Andersen 
{\it et al}~\cite{andersen}, which is at present pushed to the 
2-loop~\cite{andersen2} and 3-loop level~\cite{andersen3}. However, 
the 1-loop HTLpt pressure has a bad perturbative LO content, in the sense 
of severe over-inclusion of the effect of order $g^2$ as the HTL action 
is accurate only for soft momenta and for hard ones only in the vicinity 
of light cone. Such problem is cured (or at least pushed to higher orders) 
only after going to 2-loop level in HTlpt~\cite{andersen2}, which is indeed
a considerably more involved calculation. A very recent 1-loop QNS 
calculation~\cite{jiang} from pressure in HTLpt~\cite{andersen} 
had, obviously, the problem of over-inclusion of the order $g^2$. 
Moreover, it required an ad hoc separation scale to distinguish between 
hard and soft momenta. On the other hand the calculation of QNS in 
Ref.~\cite{munshi} dealt with the imaginary part of the charge-charge 
correlator in the vector channel and required 
to show the charge conservation. It also encountered some technical 
difficulties and over-inclusion problem in order $g^2$ as discussed in 
Ref.~\cite{blaizot}. Also the Landau damping (LD) contribution was discussed 
but ignored. 

The equation of state (EOS) of strongly interacting matter at nonzero baryon 
density and high temperature is a subject of great interest for wide 
spectrum of physicists. Also QNS is a topical quantity in 
view of the ongoing efforts towards understanding the actual nature 
of the QGP~\cite{white} as QNS plays an important
role~\cite{stephanov,jeon} in locating the critical end point in 
QCD phase diagram.  As it stands the LO thermodynamic 
quantities~\cite{andersen,blaizot1} and QNS~\cite{blaizot,munshi,jiang} 
in HTL approximation~\cite{braaten} led to different results. This
requires a detailed analysis of the leading order
quantities within the HTLpt before 
extending it to the higher orders. In view of this we do not aim at higher 
orders calculations, rather {\it we intend in this article to sort out 
the problems in 1-loop HTLpt}, which produced different results from that 
of the $\Phi$-derivable approach within the HTL resummation~\cite{braaten}, 
and finally arrive at {\it a consistent result despite the use of different 
approaches.} 

The paper is organised as follows. In Sec. II we briefly discuss some
generalities on fluctuations, correlation functions and susceptibilities
based on external disturbance to a physical system. In Sec. III  we modify
the HTLpt based on the external disturbance. In Sec. IV the HTL 
thermodynamics in presence of the quark chemical potential and then 
QNS in LO are obtained.  We also checked, both 
numerically and analytically, the  perturbative content of LO HTL QNS in 
Sec. V.  Finally, we conclude in Sec. VI.

\section{Generalities}
 \subsection{ Fluctuation and Susceptibility:}

Let \( {\cal O}_{\alpha } \) be a Heisenberg operator where $\alpha$ may be
associated with a degree of freedom in the system. In a static and uniform
external field \( {\cal F}_{\alpha } \), the (induced)
expectation value of the operator \( {\cal O}_\alpha \left( 0,\overrightarrow{x}
\right) \) is written~\cite{kunihiro} as
\begin{equation}
\phi _{\alpha }\equiv \left\langle {\cal O} _{\alpha }\left
( 0,\overrightarrow{x}\right) \right\rangle _{\cal F}=
\frac{{\rm Tr}\left
[ {\cal O} _{\alpha }\left( 0,\overrightarrow{x}\right) e^{-\beta \left
( {\cal H}+{\cal H}_{ex}\right) }\right] }{{\rm Tr}\left[ e^{-\beta
\left( {\cal H}+{\cal H}_{ex}\right) }
\right] }=\frac{1}{V}\int d^{3}x\, \left\langle {\cal O} _{\alpha }
\left( 0,\overrightarrow{x}\right) \right\rangle \: , \label{eq1}
\end{equation}
where the translational invariance is assumed, $V$ is the volume of the
system and
\({\cal H}_{ex} \) is given by
\begin{equation}
{\cal H}_{ex}=-\sum _{\alpha }\int d^{3}x\, {\cal O} _{\alpha }\left( 0,
\overrightarrow{x}\right) {\cal F}_{\alpha }\: .\label{eq2}
\end{equation}

The (static) susceptibility \( \chi _{\alpha \sigma } \) is defined as
the rate with which the expectation value changes
in response to an infinitesimal change in external field,
\begin{eqnarray}
\chi _{\alpha \sigma }(T) & = & \left. \frac{\partial \phi _{\alpha }}
{\partial {\cal F}_{\sigma }}\right| _{{\cal F}=0}
  = \beta \int d^{3}x\, \left\langle {\cal O} _{\alpha }\left
( 0,\overrightarrow{x}\right) {\cal O} _{\sigma }( 0,\overrightarrow{0})
 \right\rangle \: , \label{eq3}
\end{eqnarray}
where
$\langle {\cal O}_\alpha (0,{\vec x}){\cal O}_\sigma(0,{\vec 0})\rangle $
is the two point correlation function with operators evaluated
at equal times. There is no broken symmetry as
\begin{equation}
\left.\left\langle {\cal O} _{\alpha }
\left ( 0,\overrightarrow{x}\right ) \right\rangle
\right|_{{\mathcal F}\rightarrow 0}
 =\left. \left\langle {\cal O} _{\sigma } 
( 0,\overrightarrow{0}) \right\rangle 
\right |_{{\mathcal F}\rightarrow 0}=0  \ . \label{eq3i}
\end{equation}

\subsection{{Thermodynamic Relations:}}
The pressure is defined as
\begin{equation}
{\cal P}=\frac{T}{V} \ln {\mathcal Z}\ , \label{def_press} 
\end{equation}
where $T$ is temperature, $V$ is the volume and 
${\mathcal Z}$ is the partition function of a quark-antiquark gas.  
The entropy density is defined as
\begin{equation}
{\cal S}=\frac{\partial {\cal P}}
{\partial T} \ . \label{def_entropy}
\end{equation}
The  number density for a given 
quark flavour can be written  as
\begin{equation}
\rho = \frac{\partial {\cal P}} {\partial \mu} 
=\frac{1}{V}
\frac{{\rm{Tr}}\left [  {\cal N} e^{-\beta \left({\cal H}-\mu {\cal N}
\right )}\right ]}
{{\rm{Tr}}\left [e^{-\beta \left({\cal H}-\mu {\cal N}\right )}\right ]} =
\frac{\langle {\cal N}\rangle}{V} 
\ , \label{eq5}
\end{equation}
with  ${\cal N}$ is the quark number operator and $\mu$ is the chemical 
potential. If $\mu\rightarrow 0$, 
the quark number density vanishes due to CP invariance.

The QNS is a measure of the response of the quark number density with 
infinitesimal change in the quark chemical potential, $\mu+\delta\mu $,
at $\mu\rightarrow 0$. Under such a situation the variation of the
external field, ${\cal F}_\alpha$, in ({\ref{eq2}}) can be identified 
as the quark chemical potential $\mu$ and the operator ${\cal O}_\alpha$ as
the temporal component ($J_0$) of the external vector current, 
$J_\sigma(t,{\vec x})= \overline{\psi} \Gamma _{ \sigma}\psi$, 
where $\Gamma_\sigma$ is in general a three point function.
Then the QNS for a given quark flavour follows from
(\ref{eq3}) as
\begin{eqnarray}
\chi(T) &=& \left.\frac{\partial \rho}{\partial \mu}\right |_{\mu=0}
= \left.\frac{\partial^2 {\cal P}}{\partial \mu^2}\right |_{\mu=0}
= \int \ d^4x \ \left \langle J_0(0,{\vec x})J_0(0,{\vec 0})
\right \rangle \ =- \lim_{p\rightarrow 0} {\mbox{Re}} \Pi^R_{00}(0,p),
\label{eq4}
\end{eqnarray}
where the number operator,  
 ${\cal N}=\int J_0(t, {\vec x}) \ d^3x =
\int {\bar \psi}(x)\Gamma_0\psi(x)d^3x$ and
$\Pi^R_{00}({\omega_p,p })$ is the retarded time-time component 
of the Fourier transformed vector correlator 
$\Pi_{\sigma\nu}(\omega_p,{\vec p})$
with an external momentum $P=(\omega_p,|{\vec{\mathbf p}}|=p)$. To write
(\ref{eq4}) in such a compact form we have used the fluctuation-dissipation 
theorem and the quark number conservation~\cite{kunihiro,calen}, 
$\lim_{{\vec p}\rightarrow 0}{\rm{Im} \Pi^R_{00}(\omega_p,p)}\propto 
\delta(\omega_p)$. 

\section{Modification on Hard Thermal Loop Perturbation Theory}
The HTL Lagrangian density for quark including HTL correction 
term~\cite{pisarski} is written as 
\begin{eqnarray}
\mathcal{L}_{HTL}&=&\mathcal{L}_{QCD}+\delta{\mathcal{L}}_{HTL}
\nonumber\\
&=&\bar\psi i\gamma_\mu D^\mu \psi
+m_q^2\bar\psi\gamma_\mu\left
\langle \frac{R^\mu}{iR\cdot D}\right\rangle\psi \ , \label{i1}
\end{eqnarray}
where $R=(1,{\mathbf r})$ is a light like four-vector, $\psi$ and 
$\bar{\psi}$ are the fermionic fields, $D$ is the covariant derivative, 
$\langle\rangle$ is the average over all possible directions over loop
momenta.  The second term is gauge invariant, nonlocal and can generate 
$N$-point HTL functions~\cite{braaten}, which are inter-related through 
Ward identities. Now $m_q$, is the quark mass in a hot and dense medium, 
which depends on the strong coupling $g$, temperature $T$ and the chemical
potential $\mu$. Despite these facts $m_q$ is treated in (\ref{i1}) as a 
parameter much like the rest mass of a quark and a HTLpt has been 
developed~\cite{andersen} around this rest mass ({\em i.e.}, $m_q^2$)  
by reorganising the HTL terms where $m_q^2$ was treated as the order 
of $(gT)^0$ for a hot system. In this way the effect of $m_q^2$ is taken 
into account in higher orders much like a variational principle. For a hot
and dense system we will also treat the mass parameter as the order of
$(gT)^0$ and $(g\mu)^0$, and reorganise the HTL term based on the variation
of the external source and setting it zero at the end. 
In this way the effect of $m_q^2$ is taken into account to the higher order 
variations of the external source and thus to the response of the system. 

We note that the covariant derivative usually contains background field
or any source, $j$ depending upon the physical requirement. To motivate 
the perspective we define the covariant derivative $D^\mu$ as
\begin{equation}
D^\mu= [{\cal D}^\mu-i\delta^{\mu 0}(j+\delta j)]= 
[{\tilde D}^\mu -i\delta^{\mu 0}\ \delta j] \ . \label{i1c}
\end{equation} 
We note that ${\cal D}$ contains gauge coupling and 
 ${\tilde D}^\mu= {\cal D}^\mu-i\delta^{\mu 0}j$, and  $\delta j$ 
is an infinitismal change in external source to which the 
response of the system can be calculated, as discussed in Sec.II. 
Later it can be identified with a variation of some physical quantity 
depending upon the requirement of the system
under consideration.  

Now, expanding the second term in (\ref{i1}), we can write as
\begin{eqnarray}
\mathcal{L}_{HTL}(j+\delta j) &=& \bar\psi\left(i\Slash\! {\tilde D} 
+ \Sigma\right) \psi
+ \delta j \ \bar\psi \Gamma_0 \psi
+  \delta j^2\ \bar\psi \frac{\Gamma_{00}}{2}\psi
+ {\mathcal O}(\delta j^3)\ \nonumber \\
&=&\mathcal{L}_{HTL}(j) 
+ \delta j \ \bar\psi \Gamma_0 \psi
+  \delta j^2\ \bar\psi \frac{\Gamma_{00}}{2}\psi
+ {\mathcal O}(\delta j^3)\ 
\label{i2}
\end{eqnarray}
where the various $N$-point functions in coordinate space are generated as
\begin{eqnarray}
\Sigma &=& m_q^2\left
\langle \frac{\Slash\! R}{i R\cdot {\tilde D}}\right\rangle , \   
\Gamma_0= 
\delta^{\mu 0}\gamma_\mu - m_q^2
\left\langle \frac{\Slash\!RR_\mu\delta^{\mu 0}}{(iR\cdot {\tilde D})^2}
\right\rangle , \  
\Gamma_{00}=  2 m_q^2 \bar\psi\left\langle
\frac{\Slash\! R R_\mu R_\nu\delta^{\mu 0}\delta^{\nu 0}}  
{(iR\cdot {\tilde D})^3} \right\rangle , 
\label{i2n}
\end{eqnarray}
where these functions can easily be transformed into momentum 
space~\cite{lebellac}. 
We now note that these $N$-point HTL functions in (\ref{i2n}) are 
also inter-related by Ward identities.  In 
HTL-approximation the 2-point function, $\Sigma \ \sim \ gT$ (quark-self
energy) is of the same order as the tree level one, $S_0^{-1}(K) \sim 
K\!\!\!\! \slash \ \sim gT$ (in the weak coupling limit $g< < 1$), if the 
external momenta are soft, {\it i.e.}, of the order of $gT$. The 3-point
function is given by $g\Gamma_\nu=g(\gamma_\nu+\delta \Gamma_\nu)$, where
$\delta\Gamma_\nu$ is the HTL correction. The 4-point function,
$g^2\Gamma_{\nu\sigma}$, is higher order and  does not exist at the bare
perturbation theory 
and only appears within the HTL approximation~\cite{pisarski,braaten}. 

Now considering the HTL Lagrangian in (\ref{i2}), we can write the 
partition function~\cite{lebellac} as
\begin{eqnarray}
{\cal Z}[\beta;j+\delta j]=\int {\cal D}[\bar \psi] 
{\cal D}[{\psi}] {\cal D}[{\cal A}]
e^{i\int d^4x\mathcal{L}_{HTL}(\psi,{\bar \psi};j+\delta j)},
\label{i3}
\end{eqnarray}
where $\beta=1/T$, is the inverse of the temperature and ${\cal A}$ is a 
background gauge field.

The pressure can be written as
\begin{equation}
{\cal P}[\beta:j+\delta j]=\frac{1}{\cal V} \ln{\cal Z}[\beta:j+\delta j]
\ , \label{i3_1}
\end{equation}
where the four-volume, ${\cal V}=\beta V $ with 
$V$ is the three-volume.

Expanding ${\cal P}$ in Taylor series around  $\delta j$ one can write
\begin{eqnarray}
{\cal P}[\beta:j+\delta j] &=& {\cal P}[\beta;j]+ \delta j \ 
\left. {\cal P}'[\beta;j+\delta j]\right |_{\delta j\rightarrow 0}
+\frac{{\delta j}^2}{2}\ \left. {\cal P}''[\beta;j+\delta j]
\right|_{\delta j\rightarrow 0}
+ \cdots \cdots \ .
\label{i3_1t}
\end{eqnarray}
The first derivative of ${\cal P}$ w.r.t. $j$ is related to the 
conserved density in (\ref{eq1}) whereas the second derivative is related
to the conserved density fluctuation in (\ref{eq3}). The above expansion
in (\ref{i3_1t}) is very important for a resummed perturbation theory.
We now note that a HTLpt was developed in Ref.~\cite{andersen}
by considering the first term in (\ref{i3_1t}) with $j=0$, which caused 
an over-inclusion of the LO pressure. This was cured
by going into {\it two-loop} level in HTLpt~\cite{andersen2}, which is
of-course a very involved in nature.  As we will see below this 
could easily be corrected if one constructs a HTLpt at the first 
derivative level of ${\cal P}$ in (\ref{i3_1t}) where the effect of the 
variation of external field is taken into account.  

Now ${\cal P}'$ can be obtained as
\begin{eqnarray}
\!\!\!\!\!
\!\!\!\!\!
\left. \frac{\del{\cal P}[\beta; j+\delta j]}{\del j}
\right |_{\delta j\rightarrow 0}
\!\!\!\!\!
&=&\frac{i}{{\cal V}{\cal Z}[\beta;j]}\ {\int {\cal D}[\bar \psi] 
{\cal D}[\psi] {\cal D}[{\cal A}]
\int d^4x \ {\bar \psi(x)} \Gamma_0[j] \psi (x)} \,
e^{ \left ({i\int d^4x {\cal L}_{HTL}
(\psi,{\bar{\psi}}; j)}\right)} \ ,
\label{i3_2} 
\end{eqnarray}
where we have used (\ref{i2n}). The full HTL quark propagator in presence 
of uniform $j$ can be written as
\begin{eqnarray}
{\cal S}_{\alpha\sigma}[j](x,x')&=&\frac{
\int {\cal D}[\bar \psi] {\cal D}[\psi] {\cal D}[{\cal A}]
\psi_{\alpha}(x){\bar \psi_\sigma(x')} \exp \left ({i\int d^4x {\cal L}_{HTL}
(\psi,{\bar{\psi}}; j)}\right)}
{ \int {\cal D}[\bar \psi] {\cal D}[\psi] {\cal D}[{\cal A}]
 \exp \left ({i\int d^4x {\cal L}_{HTL} 
(\psi,{\bar{\psi}};j)}\right)} \ .
\label{i3_fp}
\end{eqnarray}
We now note that this full HTL propagator, ${\cal S}[j]$, is indeed difficult 
to calculate and we would approximate it by 1-loop HTL resummed
propagator~\cite{pisarski,braaten}, $S^\star[j]$ and also other HTL 
functions below.

Now using (\ref{i3_fp})  and performing the traces over the colour, flavour,
Dirac and coordinate indices in (\ref{i3_2}) one can write
\begin{eqnarray}
\left.\frac{\del{\cal P}[\beta;j+\delta j]}{\del j}\right |_{\delta j=0}
&=& \ -i\!\!\int\! \frac{d^4K}{(2\pi)^4} 
\mbox{tr}\left [S^\star[j](K)\ \Gamma_0[j](K,-K;0) \right ] , 
\label{i3_tr}
\end{eqnarray}
where '{\rm{tr}}' indicates the trace over the colour, flavour and 
Dirac indices.

Similarly, we obtain ${\cal P}''$ as 
\begin{eqnarray}
\left.\frac{\del^2{\cal P}[\beta;j+\delta j]}{\del j^2}
\right|_{\delta j\rightarrow 0} 
&=& -N_cN_fT\!\! \int \frac{d^3k}{(2\pi)^3}\nonumber \\
&\times& \sum_{k_0}\mbox{Tr}
\left [S^\star[j](K)\ \Gamma_0[j](K,-K;0) \ S^\star[j](-K)\ \Gamma_0[j](K,-K;0) 
\right. 
\nonumber \\
&&\left.  
- S^\star[j](K)\  \Gamma_{00}[j](K,-K;0,0) 
\right ] \, , 
\label{i3_d2}
\end{eqnarray}
where $N_f$ is the number of massless flavours,
$N_c$ is the number of colour and '{\mbox{Tr}}' indicates the trace over only
the Dirac matrices. We have also used an identity based on unitarity of 
$S^\star[j]$ as
\begin{eqnarray}
\!\!\!\!\!\!
\!\!\!\!\!\!
\frac{\del S^\star[j](K)}{\del j}
&=&  -S^\star[j](K)\ \frac{\del {S^\star}^{-1}[j](K)}
{\del j} \ S^\star[j](K)
= -S^\star[j](K)\ \Gamma_0[j](K,-K;0)\ S^\star[j](K)
\ . \label{s0_ii} 
\end{eqnarray}

Now, if we identify $j$ as the quark chemical potential $\mu$, and 
$\delta j$ as its change 
$\delta\mu$, then (\ref{i3_d2}) would represent the QNS as
\begin{eqnarray}
\chi(\beta)\!\!&=&\!\!\!
\left.\frac{\del{\rho}}{\del \mu}
\right|_{\mu \rightarrow 0}
=\left.\frac{\del^2{\cal P}}{\del \mu^2}
\right|_{\mu \rightarrow 0} 
= - N_cN_fT\!\! \int \frac{d^3k}{(2\pi)^3} \nonumber \\  
&\times&\!\!\!\!\!\!\!\!\!\!\!\!\!\! \sum_{k_0=(2n+1)\pi i T}\!\!\!\!\!\!\!\!\!
\mbox{Tr} \left [S^\star (K)  \Gamma_0(K,-K;0)S^\star(K)\Gamma_0(K,-K;0)
 - S^\star (K) \Gamma_{00}(K,-K;0,0)\right ], 
\label{s1}
\end{eqnarray}
where the temporal correlation functions at the external momentum
$P=(\omega_p,|{\vec p}|)=0$, is related to the thermodynamic 
derivatives\footnote{As already discussed Ref.~\cite{munshi} dealt 
with the definition of QNS that involves the static limit of the imaginary
part of the dynamical charge-charge correlator. If one uses the number
conservation directly, {\it viz.}, the imaginary part of the charge-charge
correlator is proportional to $\delta(\omega_p)$, then it becomes equal to
(\ref{s1}) in which charge conservation is in-built by construction.}.
The first term in the second line of (\ref{s1}) is a 1-loop self-energy
whereas the second term corresponds to a tadpole in HTLpt with effective
$N$-point HTL functions. 


Now, (\ref{i3_tr}) represents the  LO net quark number density in 
presence of uniform external field $\mu$ as 
\begin{eqnarray}
\rho(\beta,\mu)&=& \frac{\del{\cal P}}
{\del \mu} 
= \ -i\!\!\int\! \frac{d^4K}{(2\pi)^4} 
\mbox{tr}\left [S^\star[\mu](K)\ \Gamma_0[\mu](K,-K;0) \right ] \nonumber \\ 
&=& N_cN_fT \int \frac{d^3k}{(2\pi)^3} \sum_{k_0=(2n+1)\pi i T+\mu}
\mbox{Tr} \left [S^\star(K)  \Gamma_0(K,-K;0)\right ] \ .
\label{i3_3} 
\end{eqnarray}

\begin{figure}[!tbh]
{\includegraphics[height=0.17\textwidth, width=0.5\textwidth]{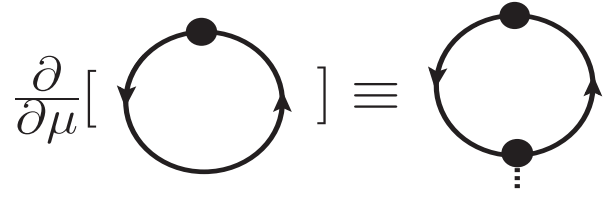}}
\caption{ 1-loop Feynman diagram in HTLpt for quark number density, $\rho_q$ 
that originates with the variation of $\mu$ of the 1-loop HTLpt pressure.
The dashed line represents the background field. The solid blobs are 1-loop
resummed HTL $N$-point functions.}
\label{rho_htl}
\end{figure}

\begin{figure}[!tbh]
{\includegraphics[height=0.17\textwidth, width=0.5\textwidth]{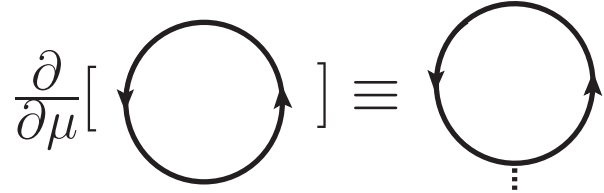}}
\caption{ Same as Fig.\ref{rho_htl} but for the lowest order bare 
perturbation theory.}
\label{rho_f}
\end{figure}

\begin{figure}[!tbh]
{\includegraphics[height=0.17\textwidth, width=0.5\textwidth]{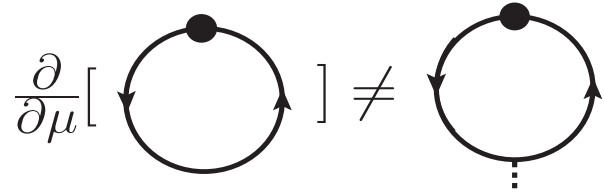}}
\caption{ The $\ne$ sign indicates that it is not the correct diagram in 
the right hand side (rhs) as the $\mu$ variation is not taken into account
properly. The diagram in rhs actually corresponds to $\rho_q$ that was 
obtained in Ref.~\cite{jiang}.}
\label{rho_ji}
\end{figure}

When $\mu\rightarrow 0$, the net quark density in (\ref{i3_3}) would
vanish as there is no broken CP symmetry, which becomes consistent with 
(\ref{eq3i}).  Also, (\ref{i3_3}) constitutes a LO HTLpt in the first 
derivative level of ${\cal P}$ (see Fig.~\ref{rho_htl}) similar to the 
usual perturbation theory where the bare $N$-point functions 
(see Fig.~\ref{rho_f}) are automatically replaced by the 1-loop resummed HTL
$N$-point functions. This suggests that HTL resummation technique provides
a consistent perturbative expansion if one goes beyond the lowest order 
perturbation theory. In contrast Ref.~\cite{jiang} did not employ the 
variation of the external source as done in (\ref{i3_1t}), which leads 
to Fig.~\ref{rho_ji} with a bare vertex for the calculation of the 
net quark number density. This resulted in overcounting of the LO QNS.  
It also required an ad hoc separation scale to distinguish between 
soft and hard momenta. 

Below we briefly outline some of the essential quantities in HTL 
resummation~\cite{braaten}, which  are required to compute (\ref{i3_3}). 
The resummed HTL propagator in 1-loop approximation 
for momentum $K$ is given as 
\begin{equation}
S^\star(K)=-\frac{\gamma_0-\vec\gamma\cdot\hat k}{2D_+(k_0,k)}-\frac{\gamma_0+
\vec\gamma\cdot\hat k}{2D_-(k_0,k)} \ , \label{H1}
\end{equation}
with
\begin{eqnarray}
 D_{\pm}(k_0,k)&=& -k_0\pm k + \frac{m_q^2}{k}\left[ \frac{1}{2}\left ( 
1\mp\frac{k_0}{k} \right ) \ln\frac{k_0+k}{k_0-k}\pm 1 \right ],
\label{H1p} \\ \nonumber\\
m_q^2 &=& \frac{g^2}{6}\left(T^2+
\frac{\mu^2}{\pi^2}\right ) \ .
\end{eqnarray}
where $g^2=4\pi\alpha_s$, $\alpha_s$ is the strong coupling. Now, the 
zeros of $D_\pm$ describe~\cite{yaun} the in-medium propagation or
{\em quasiparticle} (QP) dispersion of a particle excitation with energy
$\omega_+$ having chirality to helicity ratio $+1$, and of a mode 
called {\em plasmino} with energy $\omega_-$ having chirality to 
helicity ratio $-1$. In addition, $D_\pm$ contains a discontinuous
part corresponding to {\em Landau Damping} (LD) due to the presence
of Logarithmic term in (\ref{H1p}). Using these general properties
of the quark propagator one can obtain the in-medium spectral function 
for quarks. 

The pole part of the spectral function can be written as 
\begin{eqnarray}
\varrho_\pm(\omega,k)=\frac{\omega_\pm^2-k^2}{2m_q^2}\delta(\omega-\omega_\pm)
+\frac{\omega_\mp^2-k^2}{2m_q^2}\delta(\omega+\omega_\mp) \ , \label{i7}
\end{eqnarray}
as $D_+$ has poles at $\omega_+$ and $-\omega_-$ whereas those of
$D_-$ are at $\omega_-$ and $-\omega_+$.

For $k_0^2<k^2$, there is a discontinuity in $\ln\frac{k_0+k}{k_0-k}$
as $\ln{y}=\ln\left |y \right |-i\pi \ ,$ which leads to the spectral 
function, $\beta_\pm(\omega,k)$, corresponding 
to the discontinuity in $D_\pm(k_0,k)$ as
\begin{eqnarray}
\beta_\pm(\omega,k)\!\! &=&\!\! -\frac{1}{\pi}\mbox{Disc}\frac{1}{D_\pm(k_0,k)} 
=-\frac{1}{\pi} \mbox{Im}\left.\frac{1}{D_\pm(k_0,k)}
\right |_{k_0\rightarrow\omega+i\epsilon \atop \epsilon \rightarrow 0}
\nonumber \\
&=&\!\! \frac{\frac{ m_q^2}{2k}\left(\pm\frac{\omega}{k}-1\right)
\Theta(k^2-\omega^2)}
{\left[\omega\mp 
k-\frac{m_q^2}{k}\left(\pm 1-\frac{\omega\mp k}{2k}\ln\frac{k+\omega}{k-\omega}
\right)\right]^2+\left[\pi\frac{m_q^2}{2k}\left(1\mp\frac{\omega}{ k}\right)
\right]^2} \, . \label{L2}
\end{eqnarray}

The zero momentum limit of the 3-point HTL function can be 
obtained from the Ward identity~\cite{lebellac,kapusta} as
\begin{equation}
\Gamma^0(K,-K;0)=\frac{\del}{\del k_0} \left ({{S^\star}^{-1}(K)}\right )
=a\gamma^0+b\vec\gamma\cdot\hat k \ ,
 \label{H2}
\end{equation}
where 
\begin{equation}
a\pm b=-D_\pm'(k_0,k) , \label{H3}
\end{equation}
with 
\begin{equation}
D'_\pm = \frac{D_\pm}{k_0\mp k} - \frac{2m_q^2}{k_0^2-k^2} \ .
\label{H3p}
\end{equation}

\section{Thermodynamics and Quark Number Susceptibility}

We first obtain the net quark density $\rho(T,\mu)$, which is then
used to  obtain various thermodynamic quantities, {\em viz.},
pressure, entropy density and QNS. 

\subsection{Free case}
In free case the number density can be written from (\ref{i3_3}) as 
\begin{equation}
\rho^f(T,\mu)=N_cN_fT\int\frac{d^3k}{(2\pi)^3}\sum_{k_0=(2n+1)\pi i T+\mu}
\mbox{Tr}[S_f(K)\gamma_0], \label{f1}
\end{equation}
where the 3-point function is $\Gamma_0=\gamma_0$
and the 2-point function is the free quark propagator 
for momentum $K$ is given as
\begin{equation}
S_f(K)=-\frac{\gamma_0-\vec\gamma\cdot\hat k} {2d_+(k_0,k)}-\frac{\gamma_0+
\vec\gamma\cdot\hat k}{2d_-(k_0,k)}, \label{f2}
\end{equation}
with
\begin{equation}
d_\pm=-k_0\pm k \ \ . \label{f3}
\end{equation}
Using (\ref{f2}) in (\ref{f1}) and performing the trace over Dirac matrices, 
we get
\begin{eqnarray}
\rho^f(T,\mu)=2N_cN_fT\int\frac{d^3k}{(2\pi)^3}\sum_{k_0=(2n+1)\pi i T+\mu}
\left[\frac{1}{k_0-k}+ \frac{1}{k_0+k}\right] .\label{f4}
\end{eqnarray}
For evaluating the frequency sum in (\ref{f4}), we use the standard 
technique of contour integration~\cite{kapusta} as
\begin{equation}
\frac{1}{2\pi i}\oint\limits_C\left[\frac{1}{k_0-k}+
\frac{1}{k_0+k}\right]\frac{\beta}{2}\mbox{tanh}\left(\frac{\beta (k_0-\mu)}{2}
\right)dk_0 =\frac{\beta}{2}\ \frac{1}{2\pi i}\times 
(-2\pi i)\sum {\mbox{Residues}}\ . 
\label{f5}
\end{equation}
It can be seen that the first term of (\ref{f5}) has a simple pole at 
$k_0=k$ whereas the second term has a pole at $k_0=-k$. After calculating 
the residues of those two terms, the number density becomes
\begin{eqnarray}
\rho^f(T,\mu)&=&-N_cN_f\int \frac{d^3k}{(2\pi)^3}
\left[\tanh\frac{\beta(k-\mu)}{2}
-\tanh\frac{\beta(k+\mu)}{2}\right]
\nonumber\\
&=&2N_cN_f\int\frac{d^3k}{(2\pi)^3}\left[n(k-\mu)-n(k+\mu)\right], 
\label{f8}
\end{eqnarray}
where $n(x)=1/(e^{\beta x}+1)$, is the Fermi distribution function.

Now, the pressure is obtained by integrating the first line of (\ref{f8}) 
w.r.t. $\mu$ as
\begin{eqnarray}
{\mathcal P}^f(T,\mu) =2N_fN_c T \int \frac{d^3k}{(2\pi)^3} \left [ 
\beta k + \ln\left(1+e^{-\beta(k-\mu)}\right) 
+ \ln\left(1+e^{-\beta(k+\mu)}\right) \right ], \label{fp}
\end{eqnarray}
where the first term is the
zero-point energy that generates a usual vacuum divergence~\cite{kapusta}.
The entropy density in free case can be written from pressure as
\begin{eqnarray}
S^f(T,\mu)&=&\frac{\del P^f}{\del T}=2N_cN_f\int\frac{d^3k}{(2\pi)^3}
\Big [\ln\left(1+e^{-\beta(k-\mu)}\right)
+ \ln\left(1+e^{-\beta(k+\mu)}\right)
\nonumber\\
&& \left. 
+\frac{\beta (k-\mu)}{e^{\beta(k-\mu)}+1}+\frac{\beta (k+\mu)}
{e^{\beta(k+\mu)}+1}\right]
=N_cN_f\left(\frac{7\pi^2T^3}{45}+\frac{\mu^2T}{3}\right) \ .
\label{fs}
\end{eqnarray}

The QNS is obtained as 
\begin{eqnarray}
\chi^f(T)&=&\left.\frac{\del}{\del\mu}\left[\rho_I^f(T,\mu)\right]
\right|_{\mu=0}
=4N_cN_f\beta\int \frac{d^3k}{(2\pi)^3}n(k)\left(1-n(k)\right)
=N_fT^2 \ . 
\label{f9}
\end{eqnarray} 

\subsection{HTLpt Case}

Using (\ref{H1}), (\ref{H2}) in (\ref{i3_3}) and then performing the 
trace over Dirac matrices, the quark number density in HTLpt becomes  
\begin{eqnarray}
\rho^{HTL}(T,\mu)&=& 2N_cN_fT\int\frac{d^3k}{(2\pi)^3}
\sum_{k_0=(2n+1)\pi i T+\mu}\left[\frac{D_+'}{D_+}+
\frac{D_-'} {D_-}\right]
\nonumber\\
&=&2N_cN_fT\int\frac{d^3k}{(2\pi)^3}\sum_{k_0}\left[\frac{1}{k_0-k}
+\frac{1}{k_0+k}
 -\frac{2m_q^2}{k_0^2-k^2}\left(\frac{1}{D_+}+\frac{1}{D_-}
\right)\right]\ . \label{H4}
\end{eqnarray}
Apart from the various poles due to QPs in (\ref{H4}) it has
LD part as $D_\pm(k_0,k)$ contain Logarithmic terms which generate 
discontinuity for $k^2_0<k^2$, as discussed earlier. Equation (\ref{H4}) 
can be decomposed in individual contribution as
\begin{equation}
\rho^{HTL}(T,\mu)=\rho^{QP}(T,\mu)+\rho^{LD}(T,\mu) \ . \label{HT}
\end{equation}

\subsubsection{Quasiparticle part (QP)}
The pole part of the number density can be written as
\begin{eqnarray}
\rho^{QP}(T,\mu)&=&2N_cN_fT\int\frac{d^3k}{(2\pi)^3}
\frac{1}{2\pi i}\oint_{C'}
\left[\frac{1}{k_0-k}+\frac{1}{k_0+k} 
-\frac{2m_q^2}{k_0^2-k^2}\left(\frac{1}{D_+}+
\frac{1}{D_-}\right)\right]
 \nonumber \\ 
&& \hspace*{6cm}
\times \ \frac{\beta}{ 2}\tanh\frac{\beta(k_0-\mu)}{2}dk_0 \ . \label{H5}
\end{eqnarray}
In general residues for various poles in third and fourth terms in (\ref{H5})
can be obtained as 
\begin{eqnarray}
\mbox{Res}\left.\left\{\frac{2m_q^2}{k_0^2-k^2}\frac{1}{D_\pm}\right\}\right|_
{k_0= \omega_\pm,-\omega_\mp}= -1 \ ; \ \ \ \ && 
\mbox{Res}\left.\left\{\frac{2m_q^2}{k_0^2-k^2}\frac{1}{D_\pm}\right\}\right|_
{ k_0=\pm k}= 1 ,   \label{HR}
\end{eqnarray}
where $D_\pm(k_0=\pm k)=\pm\frac{m_q^2}{k}$. 

\begin{enumerate}
\item
First two terms in (\ref{H5}) give the same contribution as free case
in  (\ref{f8}).

\item The third term has four simple poles at $k_0=\omega_+,-\omega_-,k,-k$. 
After performing the contour integration the third term
can be written as
\begin{eqnarray}
&&\frac{1}{2\pi i} \oint_{C'} \frac{2m_q^2}{k_0^2-k^2}\frac{1}{D_+}
\frac{\beta}{2}\tanh\frac{\beta(k_0-\mu)}{2} \ dk_0 = - \frac{\beta}{2}
\left [ - \tanh\frac{\beta(\omega_+-\mu)}{2}\right. \nonumber \\
&&\left. +  \tanh\frac{\beta(\omega_-+\mu)}{2}
+\tanh\frac{\beta(k-\mu)}{2} 
-\tanh\frac{\beta(k+\mu)}{2}\right ] 
. \label{H5_3}
\end{eqnarray}

\item
The fourth term has four simple poles at $k_0=\omega_-,-\omega_+,-k, k$. 
After performing the contour integration the fourth term
can be written as
\begin{eqnarray}
&&\frac{1}{2\pi i} \oint_{C'} \frac{2m_q^2}{k_0^2-k^2}\frac{1}{D_-}
\frac{\beta}{2}\tanh\frac{\beta(k_0-\mu)}{2} \ dk_0 = -\frac{\beta}{2}
\left [ - \tanh\frac{\beta(\omega_--\mu)}{2}\right. \nonumber \\
&&\left. +  \tanh\frac{\beta(\omega_++\mu)}{2}
+\tanh\frac{\beta(k-\mu)}{2} 
-\tanh\frac{\beta(k+\mu)}{2}\right ] 
. \label{H5_4}
\end{eqnarray}
\end{enumerate}

Using (\ref{f8}), (\ref{H5_3}) and (\ref{H5_4}) in (\ref{H5})
one can obtain the HTL quasiparticle  contributions to the quark number 
density as
\begin{eqnarray}
\rho^{QP}(T,\mu)&=&-N_cN_f\int \frac{d^3k}{(2\pi)^3}\left[
\tanh\frac{\beta (\omega_+- \mu)}{2}
+\tanh\frac{\beta(\omega_--\mu)}{2} 
- \tanh\frac{\beta(k-\mu)}{2}
\right.\nonumber \\
&&\left.\hspace*{2.7cm} 
- \tanh\frac{\beta (\omega_++ \mu)}{2}
-\tanh\frac{\beta(\omega_-+\mu)}{2}
+\tanh\frac{\beta(k+\mu)}{2}
\right]
\nonumber \\
&=&2N_cN_f\int \frac{d^3k} {(2\pi)^3}\left[
n(\omega_+-\mu)+n(\omega_--\mu)-n(k-\mu) -n(\omega_++\mu)
\right.\nonumber \\
&&\left. \hspace*{4cm}
-n(\omega_-+\mu)+n(k+\mu) 
\right ]   \ \ ,
\label{H6}
\end{eqnarray}
which agrees with that of the two-loop approximately self-consistent 
$\Phi$-derivable HTL resummation of Blaizot 
{\it et al}~\cite{blaizot,blaizot1}.

Now, the pressure is obtained by integrating the first line of (\ref{H6}) 
w.r.t. $\mu$ as
\begin{eqnarray}
{\mathcal P}^{QP}(T,\mu) &=& 
2N_fN_c T \int \frac{d^3k}{(2\pi)^3} \left [ 
\ln\left(1+e^{-\beta(\omega_+-\mu)}\right) 
 +\ln\left( \frac{1+e^{-\beta(\omega_--\mu)}} 
 {1+e^{-\beta(k-\mu)}}\right ) 
\right. \nonumber \\
&&\left.
+\ln\left(1+e^{-\beta(\omega_++\mu)}\right) 
 +\ln\left( \frac{1+e^{-\beta(\omega_-+\mu)}} 
 {1+e^{-\beta(k+\mu)}}\right)
+\beta\omega_+ + \beta (\omega_--k) \  \right ]. 
\label{H7}
\end{eqnarray}
This agrees with the form given for {\em quasiparticle} contribution  by 
Andersen {\em et al}~\cite{andersen} considering the first term\footnote{ We 
note that the expression for QP  pressure in one-loop HTLpt~\cite{andersen} 
was obtained by adding and subtracting the free gas pressure.
However, in our {\it formalism the correct LO form comes out naturally} and 
no addition and subtraction is required as in Ref.~\cite{andersen}. This is
because the fluctuation of the conserved density is appropriately taken
into consideration in the present formalism.}
in (\ref{i3_1t}) for ${\mathbf \mu=0}$.
Both quasiparticles with energies $\omega_+$ and $\omega_-$ generate 
$T$ dependent ultra-violate (UV) divergences in LO HTL pressure, which
is an artefact of 1-loop HTL approximation~\cite{andersen,blaizot,blaizot1}. 
At very high $T$, $\omega_\pm\rightarrow k$ and (\ref{H7}) reduces to free
case as obtained in (\ref{fp}). 

The corresponding HTL QP entropy density in LO can be obtained as
\begin{eqnarray}
{\cal S}^{QP}(T,\mu)&=&\frac{\del {\cal P}^{QP}_I}{\del T}
=2N_cN_f\int\frac{d^3k}{(2\pi)^3}\left[\ln
\left(1+e^{-\beta(\omega_+-\mu)} \right)
+ \ln\left(\frac{1+e^{-\beta(\omega_--\mu)}}{1+e^{-\beta(k-\mu)}}
\right)\right.
\nonumber\\
&&+\ln\left(1+e^{-\beta(\omega_++\mu)}\right)
+ \ln\left(\frac{1+e^{-\beta(\omega_-+\mu)}}{1+e^{-\beta(k+\mu)}}\right)+
\frac{\beta (\omega_+-\mu)}{e^{\beta(\omega_+-\mu)}+1}
+\frac{\beta (\omega_--\mu)}{e^{\beta(\omega_--\mu)}+1}
\nonumber\\
&&\left .
- \frac{\beta (k-\mu)}
{e^{\beta(k-\mu)}+1}+\frac{\beta (\omega_++\mu)}{e^{\beta(\omega_++\mu)}+1}
+\frac{\beta (\omega_-+\mu)}{e^{\beta(\omega_-+\mu)}+1}
-\frac{\beta (k+\mu)}{e^{\beta(k+\mu)}+1}\right] \ , \label{7s}
\end{eqnarray}
which agrees with that of the 2-loop approximately self-consistent 
$\Phi$-derivable HTL resummation of Blaizot 
{\it et al}~\cite{blaizot1}.

The QNS in LO due to HTL QP can also be obtained 
from (\ref{H6}) as
\begin{eqnarray}
\chi^{QP}(T)=\left.\frac{\del}{\del\mu}\left[\rho_I^{QP}\right]
\right|_{\mu=0}
&=& 4N_cN_f\beta\int \frac{d^3k} {(2\pi)^3}\left [
n(\omega_+)\left(1-n(\omega_+)\right) 
\ + \ n(\omega_-)\left(1-n(\omega_-)\right) \right. \nonumber \\
&&\left. \ \ \ \ \ \ \ \ \ \ \ \ \ \ \ \ \ \ \ \ \ \ \ \ \
-\ n(k)\left(1-n(k)\right)
\right ] \ , \label{H8} 
\end{eqnarray}
where the $\mu$ derivative is performed only to the explicit $\mu$
dependence. Obviously (\ref{H8}) agrees exactly with that of 
the 2-loop approximately self-consistent  $\Phi$-derivable 
HTL resummation of Blaizot {\it et al}~\cite{blaizot}.
The above thermodynamical quantities in LO due to HTL
{\em quasiparticles} with excitation energies $\omega_\pm$ are similar 
in form to those of free case but the hard and soft contributions
are clearly separated out and one does not need an ad hoc separating scale 
as used in Ref.~\cite{jiang}.

\subsubsection{Landau Damping part (LD)}
The LD part of the quark number density follows from (\ref{H4}) 
and (\ref{L2}) as
\begin{eqnarray}
\rho^{LD}(T,\mu) &=& N_cN_f\int \frac{d^3k}{(2\pi)^3}\int
\limits_{-k}^k \frac{d\omega}{2\pi}
\left(\frac{-2m_q^2}{\omega^2-k^2}\right)\pi
\left[\beta_+(\omega,k)+
\beta_-(\omega,k)\right]\tanh\frac{\beta(\omega-\mu)}{2} \ \nonumber \\
&=&  N_cN_f\int \frac{d^3k}{(2\pi)^3}\int \limits_{-k}^k 
{d\omega}
\left(\frac{2m_q^2}{\omega^2-k^2}\right)\beta_+(\omega,k)
\left[n(\omega-\mu)- n(\omega+\mu) \right ]  \ .
\label{L3}
\end{eqnarray} 

One can obtain the pressure due to LD contribution by integrating
(\ref{L3}) w.r.t. $\mu$ as
\begin{eqnarray}
{\cal P}^{LD}(T,\mu)=N_cN_fT\int\frac{d^3 k}{(2\pi)^3}\int\limits_{-k}^k d\omega\left(
\frac{2m_q^2}{\omega^2-k^2}\right)\beta_+(\omega,k)\left[\ln\left(
1+e^{-\beta(\omega-\mu)}\right)\right.
\nonumber\\
\left.+\ln\left(1+e^{-\beta(\omega+\mu)}\right)+\beta\omega\right] \ ,
\label{L3p}
\end{eqnarray}
which has UV divergence like Andersen {\em et al}~\cite{andersen} and can
be removed using the appropriate prescription therein.

The corresponding LD part of entropy density can be obtained as 
\begin{eqnarray}
{\cal S}^{LD}(T,\mu)&=&N_cN_f\int\frac{d^3 k}{(2\pi)^3}
\int\limits_{-k}^k d\omega\left(
\frac{2m_q^2}{\omega^2-k^2}\right)\beta_+(\omega,k)\left. \Big [\ln\left(
1+e^{-\beta(\omega-\mu)}\right)\right.
\nonumber\\
&&\left.+\ln\left(1+e^{-\beta(\omega+\mu)}\right)+\frac{\beta(\omega-\mu)}
{e^{\beta(\omega-\mu)}+1}+\frac{\beta(\omega+\mu)}{e^{\beta(\omega+\mu)}+1}
\right]  \ .
\label{L3s}
\end{eqnarray}

Also the LD part of the QNS becomes 
\begin{eqnarray}
\chi_I^{LD}(T)=\left.\frac{\del}{\del\mu}\left[\rho_I^{LD}(T,\mu)\right ]
\right|_{\mu=0}
&=&2N_cN_f\beta\int\frac{d^3k}{(2\pi)^3} 
\int\limits_{-k}^k d\omega\left(\frac{2m_q^2}{\omega^2-k^2}\right)\ 
\nonumber\\
&&\ \ \ \times \ \ 
\beta_+(\omega,k)\ n(\omega)\left(1-n(\omega)\right) \ , \label{L4}
\end{eqnarray}
where the $\mu$ derivative is again performed only to the explicit $\mu$
dependence. It is also to be noted that the LD contribution is of the
order of $m_q^4$. The LD contribution can not be compared with 
that of the 2-loop approximately self-consistent  $\Phi$-derivable HTL 
resummation of Blaizot {\em et al}~\cite{blaizot} as it does not 
have any closed form for the final expression. The numerical values
of both the QNS agree very well. 

It is clearly evident that the various LO thermodynamic quantities in 1-loop 
HTLpt can be obtained within this modified formalism at ease instead of 
pushing the calculation to a more involved approaches. Below we demonstrate 
the correct inclusion of the perturbative
content of the order $g^2$ to the QNS in HTLpt in a strict perturbative 
sense by comparing with the usual perturbation theory.

\section{QNS in Perturbative Leading Order ($g^2$)}
In conventional perturbation theory, for massless QCD the QNS has been 
calculated~\cite{kapusta,toimela} upto order $g^4\log(1/g)$ at $\mu=0$ as 
\begin{equation}
\frac{\chi_p}{\chi_f}=1- \frac{1}{2}\left (\frac{g}{\pi}\right )^2 +
\sqrt{1+\frac{N_f}{6}}\left(\frac{g}{\pi}\right )^3-
\frac{3}{4}\left(\frac{g}{\pi}\right )^4 \log\left(\frac{1}{g}\right)+
{\cal O}(g^4)\ . \label{qns_pert}
\end{equation}
We note that for all temperatures of relevance the series decreases 
with temperature and approaches the ideal gas value from the above, 
which is due to the convergence problem of the conventional perturbation 
series.

\begin{figure}[!tbh]
{\includegraphics[height=0.45\textwidth, width=0.48\textwidth]{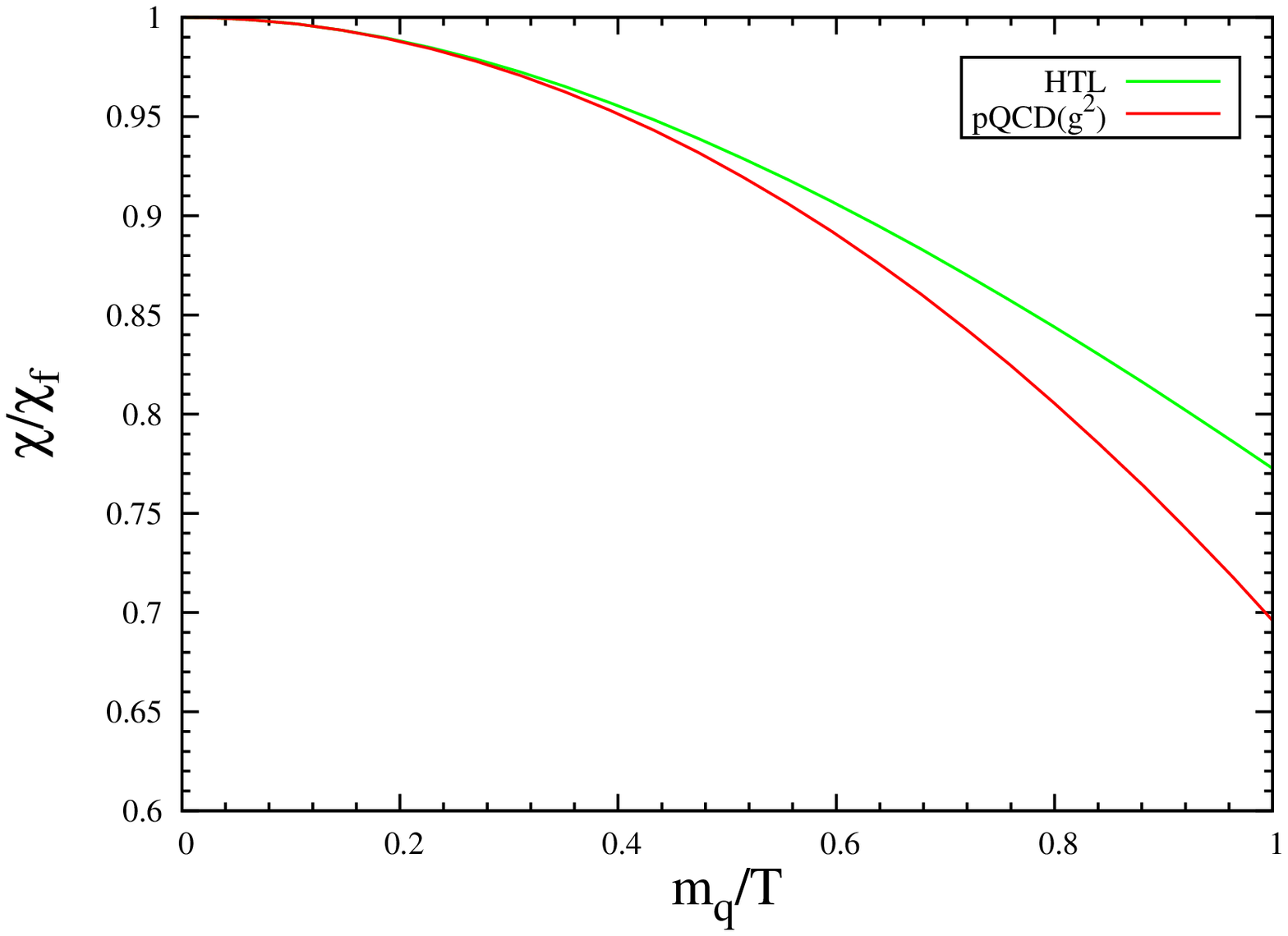}}
{\includegraphics[height=0.45\textwidth, width=0.48\textwidth]{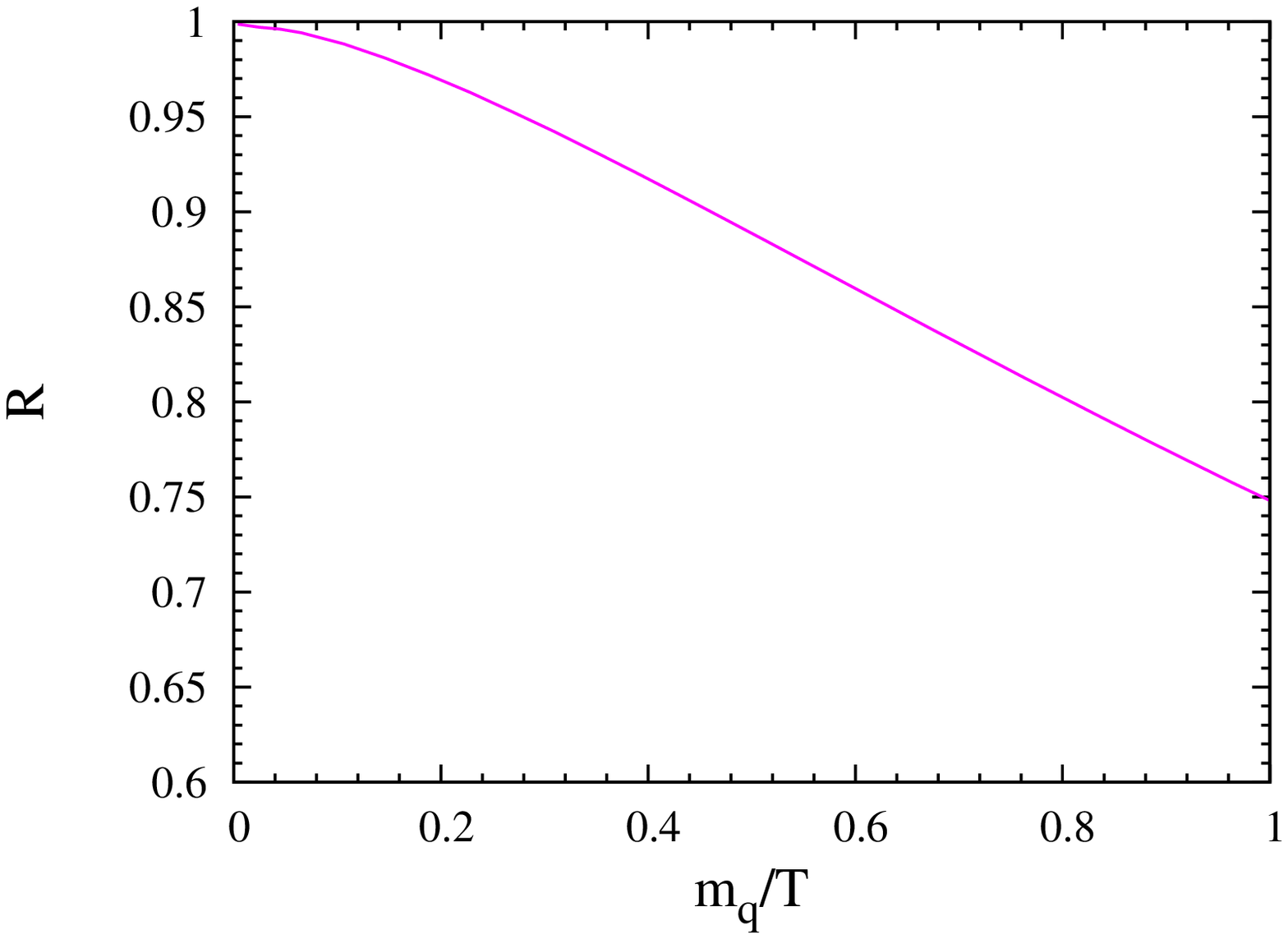}}
\caption{(Color online) {\textit {Left panel:}}The ratio of 2-flavour 
HTL to free quark
QNS and also that of LO  perturbative one  as a 
function of $m_q/T$. {\textit {Right panel:}} The interaction measure
$R$ as a function of $m_q/T$.}
\label{pert_g2}
\end{figure}

Nevertheless, the perturbative LO, $g^2$, contribution is also contained in HTL 
approximation~\cite{braaten,pisarski} through the $N$-point HTL functions. 
In the left panel of Fig.~\ref{pert_g2} we display the LO HTL QNS and LO 
perturbative QNS scaled with free one as a function of $m_q/T$, the effective
strong coupling. In the weak coupling limit both approach unity whereas the 
HTL case has a little slower deviation from the ideal gas value than the 
LO of the conventional perturbative one. The latter could be termed as an 
improvement over the conventional perturbative results. The results are in 
very good agreement with that of Ref.~\cite{blaizot}. 

Next we consider a ratio~\cite{blaizot} as
\begin{equation}
R\equiv \frac{\chi_{htl}-\chi_f}{\chi_{p(g^2)}-\chi_f} \ \ ,
\label{ratio}
\end{equation}
which measures the deviation of interaction of $\chi_{htl}$ from that of pQCD
to order $g^2$. In the right panel of Fig.~\ref{pert_g2} we display this ratio
as a function of $m_q/T$, which approaches unity in the weak coupling limit
indicating the correct inclusion~\cite{blaizot} of order $g^2$ in our approach
in a strictly perturbative sense. This comes from the $\omega_+$ 
branch~\cite{lebellac} of the HTL dispersion relations, 
$\omega_+(k)\approx k+m_q^2/k$, at hard momentum scale, {\em i.e}, $k\sim T$. 
With this one can now trivially show by expanding the QP contribution 
from (\ref{H8}) in Sect. IV that becomes
\begin{equation}
\chi_q^{QP}=\chi_f(1-g^2/2\pi^2+\cdots) , \label{pert_lo}
\end{equation}
which agrees with that in (\ref{qns_pert}). The LD contribution 
is of the order of $m_q^4$.

We note that the HTL resummation technique provides a consistent perturbative
expansion for gauge theories at finite temperature and/or density.
As discussed going beyond the lowest order bare perturbation theory for
quark number density, we use the HTL resummed propagator and quark-gluon
vertices in Fig.~\ref{rho_htl}. The resummed HTL quark propagators correspond 
to static external quarks (valence quark). In 1-loop HTLpt ({\it viz.}, 
Fig.~\ref{rho_htl}) there is no dynamical quark (no quark loop) and in this
sense 1-loop HTlpt is comparable with the quenched approximation of lattice 
QCD~\cite{haque}. The inclusion of dynamical quark loops requires one
to consider the higher-order diagrams within HTLpt in which HTL resummed
gluon propagators (containing quark loops) will show up. This could 
be taken care through (\ref{i2}) as it contains the covariant derivative
with gauge coupling and the calculation is in progress.

\section{Conclusion}
In the literature the HTL resummation have been used through various 
approaches to calculate the thermodynamic quantities and also the
response of the system, {\em viz.}, the QNS to an external perturbation, 
{\em i.e.}, the quark chemical potential. This led to different results in
LO indicating the sensitivity of the methods. 
In this paper we revisited the 
thermodynamic quantities and in particular the QNS in LO
within HTLpt to arrive at similar results within the various HTL approaches. 
For this purpose we modified the existing HTLpt~\cite{andersen} at 
the first derivative level of pressure 
by incorporating an infinitesimal variation to an external source,
{\em viz.}, the quark chemical potential that disturbs the system only 
slightly. We show that the various thermodynamic quantities and the QNS 
in LO order agree with  those of the two-loop approximately self-consistent
$\Phi$-derivable HTL resummation approach~\cite{blaizot,blaizot1} existing 
in the literature. Our calculation also shows that the soft and hard 
momenta get separated out naturally and one does not require any ad hoc 
separating scale as in \cite{jiang}.  All the thermodynamic
quantities turned out to be dependent on the chemical potential automatically
due to the method employed. We also discussed that our formalism can also be 
extended for higher-order calculations. 

\begin{acknowledgments}
We are  thankful to M. Strickland
M. H. Thoma, S. Leupold and P. Chakraborty for 
critically reading the manuscript and various useful discussions on HTLpt. 
We are also grateful to S. Majhi and P. Pal for useful help during the 
course of this work.
\end{acknowledgments}

\end{document}